\documentclass[prl,twocolumn,amsmath,amssymb,superscriptaddress]{revtex4-1}
\usepackage{graphicx}
\usepackage{dcolumn}
\usepackage{bm}
\usepackage[pdftex,colorlinks=true,citecolor=blue,urlcolor=blue,linkcolor=black]{hyperref}
\usepackage[usenames, dvipsnames]{color}
\newcommand{\ket}[1]{\vert #1 \rangle}
\begin{document}


\title{Anomalous breaking of scale invariance in a two-dimensional Fermi gas}

\author{M. Holten}
\email{mholten@physi.uni-heidelberg.de}
\affiliation{Physikalisches Institut, Ruprecht-Karls-Universit\"at Heidelberg, 69120 Heidelberg, Germany}

\author{L. Bayha}
\affiliation{Physikalisches Institut, Ruprecht-Karls-Universit\"at Heidelberg, 69120 Heidelberg, Germany}

\author{A. C. Klein}
\affiliation{Physikalisches Institut, Ruprecht-Karls-Universit\"at Heidelberg, 69120 Heidelberg, Germany}

\author{P. A. Murthy}
\affiliation{Physikalisches Institut, Ruprecht-Karls-Universit\"at Heidelberg, 69120 Heidelberg, Germany}

\author{P. M. Preiss}
\affiliation{Physikalisches Institut, Ruprecht-Karls-Universit\"at Heidelberg, 69120 Heidelberg, Germany}

\author{S. Jochim}
\affiliation{Physikalisches Institut, Ruprecht-Karls-Universit\"at Heidelberg, 69120 Heidelberg, Germany}

\date{\today}

\begin{abstract}
The frequency of the breathing mode of a two-dimensional Fermi gas with zero-range interactions in a harmonic confinement is fixed by the scale invariance of the Hamiltonian. Scale invariance is broken in the quantized theory by introducing the two-dimensional scattering length as a regulator. This is an example of a quantum anomaly in the field of ultracold atoms and leads to a shift of the frequency of the collective breathing mode of the cloud. In this work, we study this anomalous frequency shift for a two component Fermi gas in the strongly interacting regime. We measure significant upwards shifts away from the scale invariant result that show a strong interaction dependence. This observation implies that scale invariance is broken anomalously in the strongly interacting two-dimensional Fermi gas.
\end{abstract}

\pacs{Valid PACS appear here}
\maketitle



Symmetries are an indispensable ingredient to any attempt of formulating a fundamental theory of nature. Yet, it is not allways true that one can make accurate predictions about the behaviour of some complex system based on the symmetries of its Hamiltonian alone. The fundamental reason behind this is the concept of \textit{symmetry breaking} \cite{Anderson1972}. Symmetry violations often have drastic effects on the state of the system, for example when some metal breaks rotational invariance and becomes ferromagnetic.  There are three different mechanisms through which a given system may violate some of the symmetries of its Hamiltonian: \textit{explicit}, \textit{spontaneous} and \textit{anomalous} symmetry breaking \cite{Holstein2014}.


Quantum anomalies are violations of an exact symmetry of a classical action in the corresponding quantized theory \cite{Weinberg1995}. They may occur when a cut-off has to be introduced to regularize divergent terms. This regulator may explicitly break some symmetry of the theory. If this symmetry is not restored even after the cut-off is removed at the end of the renormalization procedure, the symmetry is broken \textit{anomalously}.

Quantum anomalies are ubiquitous in quantum field theories and provide, important constraints on physical gauge theories like the standard model \cite{Georgi1972,Gross1972} or on string theories \cite{Green1984,Alvarez1984}. Whereas the formalisms of explicit and spontaneous symmetry breaking are frequently applied across many fields in physics \cite{Bardeen1957,Higgs1964,Michel1980}, anomalous symmetry breaking is typically associated only with high energy physics. One exception was found in molecular physics \citep{Camblong2001,Giri2008} and here we report an observation of a quantum anomaly in the field of cold atom physics.

A particular class of anomalies, called \textit{conformal anomalies}, break the scale invariance of a theory, that is invariance of the Hamiltonian under $\bm{r} \rightarrow \lambda \bm{r}$. The most prominent examples are found in field theories like QED or QCD where the renormalized coupling constants depend on the energy scale and thus break scale invariance explicitly. In ordinary quantum mechanics the $1/r^2$- and the $\delta^2$-potential in 2D are well-known examples of conformal anomalies \cite{Holstein1993,Coon2002}.

Notably, the $\delta^2$-potential is used to model contact interactions in cold atom gases in two-dimensions as $V_\text{int} \propto \sum g_\text{0} \delta^2(r_i-r_j)$. Including the kinetic term $E_\text{kin}\propto \sum p_i^2$, the total Hamiltonian scales as $H \rightarrow H / \lambda^2 $ and it is thus scale invariant. A direct quantization of the $\delta^2$-potential gives rise to inconsistent results like a bound state with diverging energy. A renormalization procedure is required to obtain a well defined and quantized theory. To this end, the bare coupling constant $g_0$ in the Hamiltonian is replaced by a renormalized coupling $g$ and a new length scale, the 2D scattering length $a_\text{2D}$, has to be introduced \cite{Mead1991,Levinsen2015}. This additional length scale anomalously breaks the scaling symmetry of the bare Hamiltonian.

For an atom cloud trapped in an external 2D harmonic potential the scale invariance of the unregularized $\delta^2$-potential translates directly into a SO(2,1) symmetry of the full Hamiltonian \cite{Pitaevskii1997,Werner2006}. As a consequence of this symmetry, the breathing or monopole mode of the cloud follows undamped oscillations at twice the trap frequency $\omega_\text{B} = 2 \omega_\text{R}$ irrespective of the interaction strength. While it was already noted in Ref.\ \cite{Pitaevskii1997} that the required regularization of the $\delta^2$-potential leads to small deviations from this result, \citet{Olshanii2010} pointed out that this is in fact an example of a quantum anomaly that can directly be accessed via accurate frequency measurements of the breathing mode in Fermi or Bose gas experiments. The anomaly originates from the SO(2,1) symmetry of the classical action that is broken in the quantized theory. A substantial theoretical effort has been made to quantify the anomalous corrections of the breathing mode frequency $\omega_\text{B}$ both at zero \cite{Olshanii2010,Taylor2012,Hofmann2012,Gao2012} and finite \cite{Chafin2013,Mulkerin2017,Daza2018} temperature.


For two-component Fermi gases anomalous shifts up to $10\,\%$ are expected that show a strong dependence on both interaction strength and temperature. At zero temperature a rather large shift to frequencies above $2\omega_\text{R}$ is predicted \cite{Hofmann2012,Gao2012} while pertubative solutions at finite temperatures show significantly reduced shifts on the order of $1-2 \,\%$ that even go below the scale invariant value of $2\omega_\text{R}$ in the strongly interacting region \cite{Chafin2013,Mulkerin2017}.

Experimentally scale invariance has been studied with 2D Bose gases in Refs. \cite{Jackson2002,Chevy2002,Hung2010}, showing no significant symmetry violation in the weakly interacting regime. The strongly interacting regime in a 2D Fermi gas was studied in Refs. \cite{Vogt2012,Baur2013} where on the level of the experimental precision no significant deviation from the scale invariant result was observed. This was attributed to the relativly high temperatures of their system and statistical errors on the same order as the expected shifts at these temperatures \cite{Hofmann2012}.


\begin{figure}
\label{fig:1}
\includegraphics[]{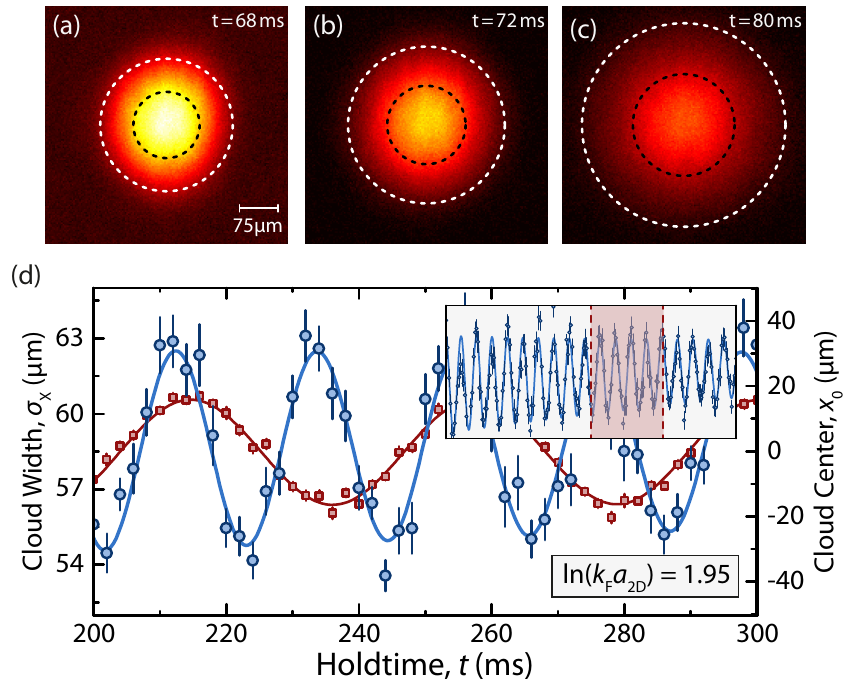}
\caption{\textbf{Breathing mode in a harmonic confinement. (a)-(c):}
In-situ images of the cloud along the strongly confined axial direction at different hold-times $t =68-80\,\text{ms}$ after quenching the trap depth at $t=0\,\text{ms}$. The inner (outer) dashed lines indicate the $1\sigma$ ($2\sigma$) width of a Gaussian fit to the cloud profile. The images are averaged over several experimental realizations and were taken at larger amplitudes to make the oscillation more apparent.
\textbf{(d):} The cloud width $\sigma_x$ as a function of the hold-time $t$ after the quench (blue circles). The inset shows the complete dataset from $t=0$ to $400\,\text{ms}$.  The center of mass oscillations of the same cloud are shown in red as a comparison (right y-axis). The solid lines are fits of a damped sine function to the measurements.}
\end{figure}

In this work, we study the anomalous frequency shift of the breathing mode in a 2D Fermi gas with high accuracy. We perform our experiments with a two-component mixture of $^6$Li atoms and approximately $10^4$ particles per spin state. The mixture is loaded into a highly anisotropic harmonic trap. The trap frequencies are $\omega_\text{z} = 2 \pi \times 7.14 \, \text{kHz}$ and $\omega_\text{R} \approx 2\pi \times 22.5 \, \text{Hz} $. The radial confinement is created by an approximately equal superposition of an optical dipole trap and a magnetic confinement proportional to $\sqrt{B}$, where $B$ is the magnetic offset field. A detailed discussion of frequency, anharmonicity and anisotropy measurements of the trap can be found in the supplementary materials \cite{som}.

The aspect ratio of approximately $300:1$ between axial and radial trap axes allows us to reach the kinematically 2D regime for low enough temperature $T$ and chemical potential $\mu$ \cite{Ries2015}. We tune the scattering length $a_\text{2D}$ by means of a broad magnetic Feshbach resonance \cite{Zuern2013}. In 2D the phase diagram of the BEC-BCS crossover is characterized by the two dimensionless parameters $\text{ln}(k_\text{F}a_\text{2D})$ and $T/T_\text{F}$ with Fermi wave vector $k_\text{F}$ and temperature $T_\text{F}$. The temperature of the cloud $T$ is extracted from the in-situ density distribution with the method established in Ref.\ \cite{Boettcher2016}. $T/T_\text{F}$ varies from $0.1$ in the BEC limit to $0.18$ in the BCS regime. The biggest effect of the quantum anomaly is expected to appear in the strongly interacting region around $\text{ln}(k_\text{F}a_\text{2D})\approx 0$ \cite{Hofmann2012}.


To excite the breathing mode, we reduce our optical confinement adiabatically such that the cloud expands in the trap. A sudden quench of the trap depth back to its original value initiates the breathing mode oscillation. By tuning the magnitude of the quench, we set the amplitude to around $8\,\%$ of the cloud width. In addition to the breathing mode, the quench leads to a small collective dipole oscillation of the center of the cloud. We do not observe any excitations of higher order collective modes in our trap using this procedure. We study both excited collective modes simultaneously by taking in-situ absorption images along the axial direction of the cloud at different times after the quench (see Fig.\ 1 (a)-(c)).

We extract the frequencies of the breathing mode $\omega_\text{B}$ and dipole mode $\omega_\text{D}$ by fitting a damped sine function to the oscillation of cloud width and center along both principal axes x and y of the trap. The principal axes of our confinement are determined and fixed by a principal component analysis of independent measurements using a non interacting gas \cite{Dubessy2014,som}. A typical dataset along the x-axis is shown in Fig.\ 1 (d). In total we obtain four frequency measurements per scattering length ($\omega_\text{B,x}$, $\omega_\text{B,y}$, $\omega_\text{D,x}$ and $\omega_\text{D,y}$).

We observe $\omega_{\text{B}} \equiv \omega_\text{B,x} = \omega_\text{B,y}$ for all interaction parameters that are accessible in our experiment. This is expected for the breathing mode in the hydrodynamic regime where the scattering rate is much larger than the oscillation frequency. The center of mass dipole modes, on the contrary, oscillate separately along both principal trap axes. From the measured difference of the two frequencies $\omega_\text{D,x}$ and $\omega_\text{D,y}$ we estimate that the in plane anisotropy of our trap is on the order of $2\,\%$ \cite{som}.

In order to compare the measured breathing mode frequency $\omega_\text{B}$ to the scale invariant result of $2\omega_\text{R}$, an accurate determination of the radial trap frequency is essential. To this end, we use the dipole frequencies that coincide with the trap frequency $\omega_\text{R}$, independent of interactions or temperature. We take the average of the two measured dipole frequencies as reference $\omega_{\text{R}} \equiv 1/2 \left( \omega_{\text{D},x} +\omega_{\text{D},y} \right)$. This is justified by the observation that the hydrodynamic breathing mode in the classical limit in a slightly anisotropic trap oscillates at this average up to a correction on the order of less than $0.1\,\%$ \cite{Merloti2013a,som}. The insensivity of the breathing mode frequency to small anisotropies is in agreement with calculations at zero temperature \cite{Hofmann2012}.


\begin{figure}\label{fig:2}
\includegraphics[]{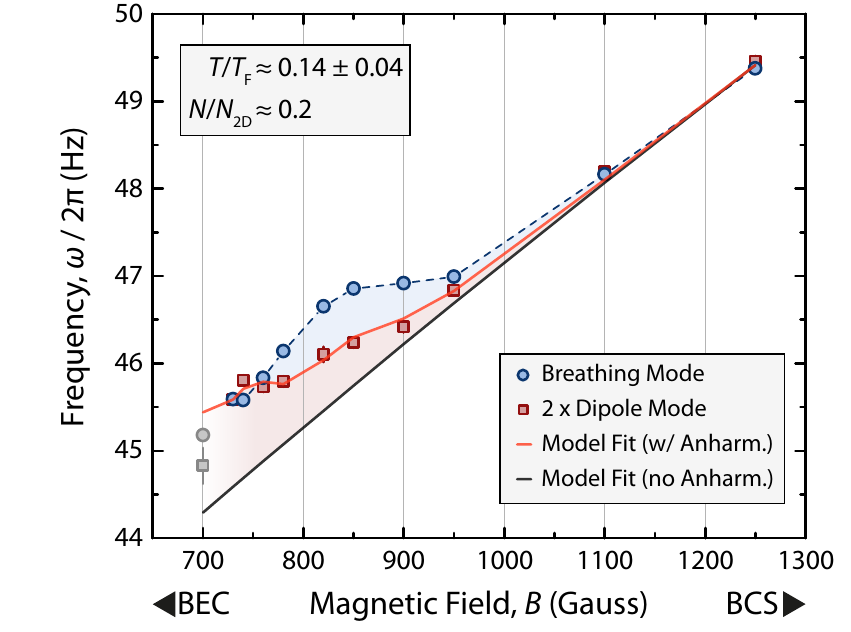}
\caption{\textbf{Measured average breathing and dipole frequencies versus magnetic offset field $\bm{B}$.} Statistical errors are on the order of the symbol size. The dipole frequencies were scaled by a factor of 2 to facilitate the comparison to the breathing mode. We fit a model $\omega_\text{R}(B, \sigma)$ for our trap frequencies to the dipole frequency measurements (solid orange line). The solid black line shows the same model for a fixed cloud size.}
\end{figure}

The measured breathing mode is very weakly damped with damping rates $\Gamma_\text{B}$ on the order of $\Gamma_\text{B} / \omega_\text{R} \approx 0.003$. The latter coincide with the background damping rate of a non-interacting cloud, confirming that, apart from technical limitations, the breathing mode is undamped. The only exception to this is the very strongly interacting region around $\text{ln}\left(k_\text{F} a_\text{2D} \right) = 0$ where we observe significantly larger, yet still small, damping rates of up to $\Gamma_\text{B} / \omega_\text{B} \approx 0.01$. This is a first indication of a broken SO(2,1) symmetry in the strongly interacting degenerate gas.

The measured average breathing and dipole frequencies as a function of the magnetic offset field are shown in Fig.\ 2. In the strongly interacting region around the Feshbach resonance at $B_0 = 832 \,\text{G}$ we find a significant shift of the breathing mode to frequencies above twice the dipole frequency (blue shaded area). In the weakly interacting BEC and BCS limits the shift disappears and the scale invariant result $\omega_\text{B} = 2 \omega_\text{D} \equiv 2 \omega_\text{R}$ is restored. The data point at $B=700\,\text{G}$ is shown greyed out due to the significant heating rates we observe this far in the BEC limit. Following \cite{Pitaevskii1997}, the observed frequency shift necessarily implies that scale invariance is broken in the strongly interacting region. As we will discuss in the following, the only conclusive explanation for the significant shift above $2\omega_\text{R}$ is the presence of the quantum anomaly. All other relevant effects which explicitly break the SO(2,1) symmetry result in a reduced breathing mode frequency instead.


To enhance our confidence in using the dipole mode measurement as reference, we fit a model for our trap frequency $\omega_\text{R}(B, \sigma )$ to the measured dipole frequencies $\omega_\text{D}$. By its dependence on the offset field $B$ and the cloud width $\sigma$ our model incorporates the magnetic field dependence and the anharmonicity of our total confinement respectively. Two free parameters of the model are determined from the fit.

\begin{figure}\label{fig:3}
\includegraphics[]{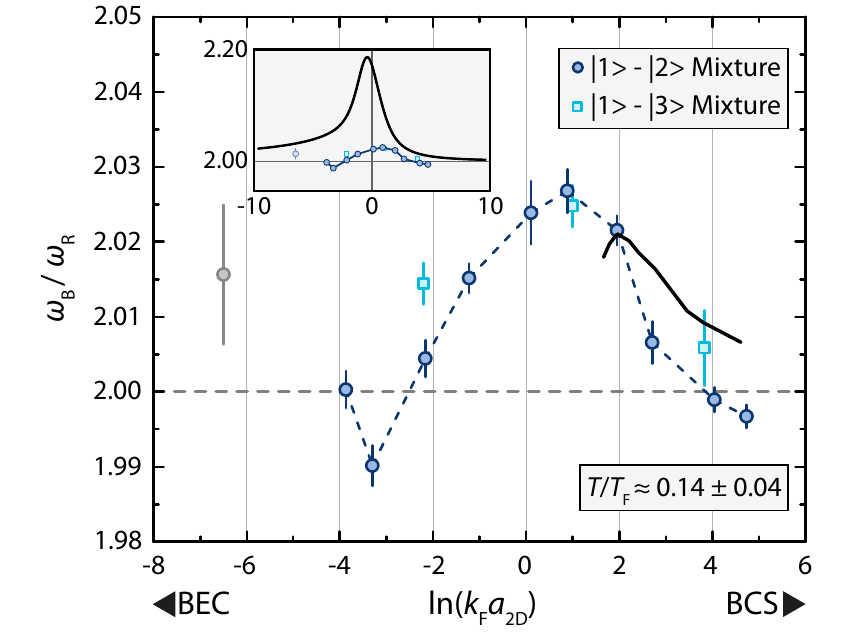}
\caption{\textbf{Anomalous shift of the breathing mode frequency.} At low temperatures the breathing mode shows a significant shift away from the scale invariant result of $\omega_\text{B} = 2\omega_\text{R}$ towards higher frequencies, even after accounting for both the effects of anharmonicity and anisotropy of our trap. Our data agrees well with a beyond mean-field approximation from Ref. \cite{Mulkerin2017} at $T/T_\text{F} =0.2$ (solid black line). The inset compares our data to a zero temperature calculation from Ref.\ \cite{Hofmann2012}. Circles and Squares represent measurement that were taken with different spin mixtures.}
\end{figure}

Our model explains the measured dipole mode frequencies remarkably well (orange solid line). We note that the origin of the visible scatter of the dipole frequencies on top of their statistical errors is just given by the fluctuations in particle numbers of different data points. These fluctuations translate into small frequency shifts through an anharmonic correction term that is proportional to the cloud width $\sigma$ squared. The overall effect of the anharmonicity can be estimated by a comparison to the same model while keeping the cloud size $\sigma_0$ fixed. We choose $\sigma_0=65 \,\mu \text{m}$ such that it matches with the measured cloud size in the BCS limit. The black solid line shows the resulting frequencies in absence of anharmonic corrections. The effective trap frequency is shifted by the anharmonicity by around $2\,\%$ in the BEC regime compared to the BCS regime (red shaded area). In the same range interactions reduce the cloud size from $\sigma=65 \,\mu \text{m}$ to $\sigma=44 \,\mu \text{m}$ in the BEC limit.

To exclude any further contributions of our trapping potential experimentally, we have performed measurements with two different spin mixtures. The difference in their Feshbach resonance positions leads to different values for $\text{ln}\left(k_\text{F} a_\text{2D}\right)$ at the same magnetic field $B$. We find no significant effect of the mixture on the measured anomalous shift (see Fig.\ 3), confirming that all magnetic field dependencies of the potential were treated properly.

As a final test, we compare the model to measurements in a non-interacting single spin component Fermi gas. Here, the anomalous frequency shift is absent and only systematic shifts from anisotropy or anharmonicity remain. Both breathing and dipole frequencies and their dependence on magnetic field and cloud width are very well explained by our model without any additional deviations and we observe no significant violation of scale invariance \cite{som}.

In Fig.\ 3 we show the relative frequency of the breathing mode $\omega_\text{B} / \omega_\text{R}$ as a function of the interaction parameter $\text{ln}\left(k_\text{F} a_\text{2D} \right)$. We observe an anomalous shift towards higher frequencies up to a maximum of $1.3\,\%$ around $\text{ln}\left(k_\text{F} a_\text{2D} \right) = 1$. The maximum position coincides with the region where we have found a many-body paired state in the normal phase of our system in a previous measurement \cite{Murthy2018} and is in agreement with zero temperature calculations \cite{Hofmann2012,Gao2012} based on a QMC simulation of the equation of state \cite{Bertaina2011}. These predict an anomalous shift of up to $10\,\%$ with a maximum at $\text{ln}\left(k_\text{F} a_\text{2D} \right) \approx -0.5$ (Fig.\ 3 inset).

The frequency shifts observed in the experiment are much smaller in magnitude. This issue is discussed extensively in literature and there are several proposed causes for the strongly reduced shift \cite{Taylor2012,Mulkerin2017,Hu2018}. Thermal fluctuations are expected to reduce the anomalous shift at finite temperatures  \cite{Hofmann2012,Gao2012,Chafin2013,Mulkerin2017} and the beyond mean field approximation from Ref. \cite{Mulkerin2017} shows anomalous frequency shifts of a similar order as our measurements at $T/T_\text{F}=0.2$ (Fig.\ 3 solid black line). Consistently, when increasing the temperature of our sample by $\Delta T = 0.1 T_\text{F}$ we observe a downwards frequency shift of the order of $-5\,\%$  \cite{som}.

In addition to the trap anharmonicity and anisotropy, we are aware of a third effect that breaks the SO(2,1) symmetry of our system explicitly: the presence of the third dimension. Fig.\ 4 shows how the third dimension affects the breathing mode frequency at fixed temperature and interactions. As we increase $N$ the quasi-2D description of our system breaks down and the system becomes kinematically three dimensional. As the system leaves the 2D limit, we observe a quick decrease of the measured frequencies below the scale invariant value of $2\omega_\text{R}$. This is in agreement with theoretical calculations which predict breathing mode frequencies of $\sqrt{10/3}\,\omega_\text{R} \left( \approx 1.83\,\omega_\text{R} \right)$ for a Bose gas and $\sqrt{3} \,\omega_\text{R}\left(\approx 1.73 \,\omega_\text{R} \right) $ for a 3D Fermi gas confined to a ``pancake'' trap in the unitary limit \cite{DeRosi2015}. Explicit breaking of scale invariance by the presence of the third dimension has been studied before both experimentally \cite{Merloti2013} and theoretically \cite{Merloti2013,Toniolo2018}. In a Bose gas a shift to lower frequencies has been observed when increasing the ratio of chemical potential to $\omega_\text{z}$.

Since both the expected and measured shifts of the breathing mode introduced by the third dimension are always negative, we conclude that measurements above $2 \omega_\text{R}$ deep in the quasi 2D limit can only be attributed to the presence of the quantum anomaly. We do however identify the third dimension as one of the possible sources for a reduced frequency measurement at any finite particle number \cite{Hu2018}. Whether the influence of finite temperature and third dimension alone explain our measurements or if additional effects, as suggested by Ref. \cite{Taylor2012}, reduce the anomalous shift further, is an interesting question to be investigated in the future.

\begin{figure}\label{fig:4}
\includegraphics[]{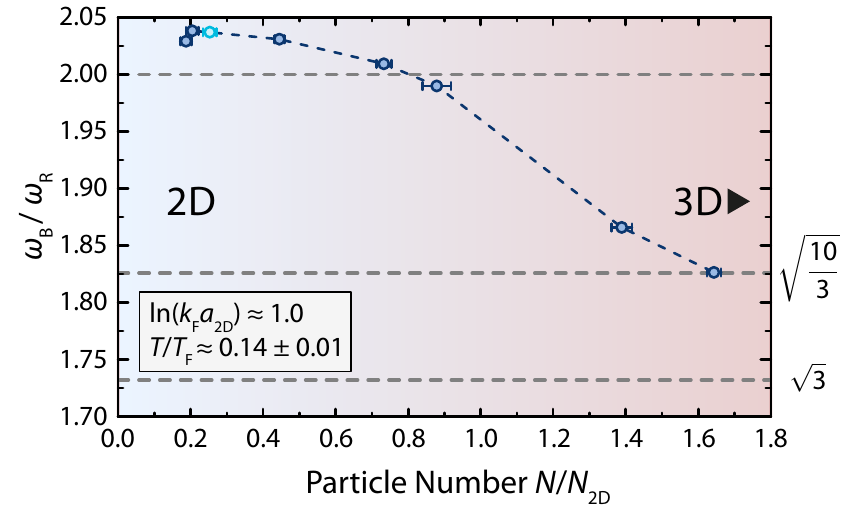}
\caption{\textbf{Influence of the third dimension on the oscillation frequency.} As we increase the particle number we observe a strong shift towards lower frequencies. On the x-axis we plot the measured particle number of one spin component $N$ divided by $N_\text{2D} = 48000$. $N_\text{2D}$ is the maximal number of non-interacting atoms per spin state in the axial ground state of our trap at this magnetic field. The dashed line is a straight connection between the data points.}
\end{figure}

The $\delta^2$-potential that we introduced as model for contact interactions is just an approximation of the actual scattering between cold atoms in nature. The exact scaling symmetry holds solely in this approximate theory. A more fundamental theory would contain a modified interaction term and the resulting Hamiltonian would break the SO(2,1) symmetry explicitly without requiring any renormalization procedure. In this equivalent picture, the same frequency shift of the breathing mode is then merely the consequence of the explicit violation of scale invariance by the Hamiltonian.

Any anomaly can be understood in this way. In the Standard Model, for instance, the appearance of quantum anomalies is related to the fact that the underlying field theories fail to accurately describe nature at small length scales. In contrast to our system, the fundamental physical description is still unknown in this case. This example highlights the significance of the concept of quantum anomalies in the formulation of effective theories that accurately describe physics at larger length scales.

To conclude, we observe a significant, interaction dependent, frequency shift away from the scale invariant frequency $\omega_\text{B} = 2\omega_\text{R}$. We have confirmed that other terms that explicitly break the symmetry of the Hamilton cannot explain the positive frequency shift of the breathing mode and we attribute it to the presence of a quantum anomaly. We have identified both temperature and the third dimension as causes of the strongly reduced anomalous shift compared to zero temperature calculations.

\hfill\\
\begin{acknowledgments}
We gratefully acknowledge insightful discussions with Tilman Enss, Nicol\'{o} Defenu, Chris Vale, H\'{e}l\`{e}ne Perrin, Maxim Olshanii and Johannes Hofmann. We thank Ralf Klemt for his help with data taking. We are aware of a study with comparable results that has been performed by the research group of Chris Vale at the University of Swinburne simultaneously to our work. This work has been supported by the ERC consolidator grant 725636, the Heidelberg Center for Quantum Dynamics and is part of the DFG Collaborative Research Centre SFB 1225 (ISOQUANT). P.M.P. acknowledges funding from European Unions Horizon 2020 programme under the Marie Sklodowska-Curie grant agreement No. 706487.
\end{acknowledgments}
\providecommand{\noopsort}[1]{}\providecommand{\singleletter}[1]{#1}%
%

\cleardoublepage
\setcounter{figure}{0}
\renewcommand{\figurename}{Fig.\! S \!\!\!}

\section*{\large SUPPLEMENTAL MATERIAL}

\section{Preparation of the Sample}
A detailed overview of the experimental setup and the preparation of our sample is found in Ref.\ \cite{Ries2015}. We are able to prepare arbitrary degenerate mixtures of the lowest three electronic hyperfine states of $^6$Li in the quasi 2D regime. We label these from lowest to highest energy as states $\ket{1}$, $\ket{2}$ and $\ket{3}$.

The experimental sequence begins by loading a $\ket{1}$-$\ket{2}$ mixture from a magneto-optical trap into a far red detuned crossed-beam optical dipole trap (ODT). In the ODT degeneracy is achieved by evaporative cooling. After evaporation, we use resonant radio frequency pulses to create balanced samples of two hyperfine states of our choice. To prepare a non-interacting single component sample, we remove one of the hyperfine states by shining in a beam of resonant light at a magnetic field of $B=1100\,\text{G}$ where the interactions are weak.

After reaching the final temperature, the sample is transferred into a single layer of a highly anisotropic standing-wave optical dipole trap (SWT) with axial and radial trap frequencies of $\omega_\text{z} = 2 \pi \times 7.14 \, \text{kHz}$ and $\omega_\text{R} \approx 2\pi \times 18.5 \, \text{Hz} $. The latter is created by intersecting two Gaussian beams under an angle of about $14^\circ$. A final spilling stage of the atoms in the SWT is utilized to set the atom number $N$ and the final temperature $T/T_f$ such that we reach the quasi-2D regime in this trap.

To tune the scattering length $a_\text{2D}$ we make use of the broad Feshbach resonances of $^6$Li. The coils that create the magnetic offset field $B$ are aligned such that they create an additional harmonic confinement in radial direction. The confinement strength is proportional to the magnetic field $B$ with trap frequencies on the order of $12$-$15\,\text{Hz}$. In combination with the SWT this leads to radial trap frequencies on the order of $\omega_\text{R}\approx22$-$24\,\text{Hz}$, depending on the magnetic offset field $B$.

We excite the breathing and dipole modes in the SWT as explained in the main text. After some hold-time $t$ we use destructive absorption imaging to obtain a single measurement of the column density of the cloud along the tightly confined direction and the sequence is repeated for a different hold-time. To reduce statistical errors we take around 20 images per hold-time. We scan the hold-time from $0$ to $400\,\text{ms}$ in steps of $2\,\text{ms}$. One experimental cycle takes around $15\,\text{s}$ leading to a non-stop measurement time of around 180 hours for the main dataset shown in Fig.\ 2 and 3.

\section{Experimental Parameters}
The phase diagram of the 2D BEC-BCS crossover is characterized by the two dimensionless parameters $T/T_\text{F}$ and $\text{ln}\left(k_\text{F} a_\text{2D} \right)$. We compute the Fermi momentum $k_\text{F}$ directly from the measured average single component density $n$ via $k_\text{F} = \sqrt{4\pi n}$. The Fermi temperature $T_\text{F}$ is related to the Fermi momentum $k_\text{F}$ as $T_\text{F} = \frac{\hbar^2 k_\text{F}^2}{2mk_\text{B}}$, where m is the mass of $^6$Li. We obtain the 2D scattering length $a_\text{2D}$ via
\begin{equation}
	a_\text{2D}=l_z \sqrt{\dfrac{\pi}{0.905}}\text{exp}\left( - \sqrt{\dfrac{\pi}{2}} \dfrac{l_z}{a_\text{3D}} \right),
\end{equation}
where $l_z = \sqrt{\hbar / m \omega_z }$ is the harmonic oscillator length in axial direction and $a_\text{3D}$ the 3D scattering length.

The absolute temperature $T$ of the gas is extracted by a fit of two reference equations of state (EOS) to the in-situ density profile. The exact form of the reference EOS and how they are used for thermometry is discussed in detail in Ref.\ \cite{Boettcher2016}.

\section{Principal Trap Axes}

\begin{figure}\label{fig:s6}
\includegraphics[]{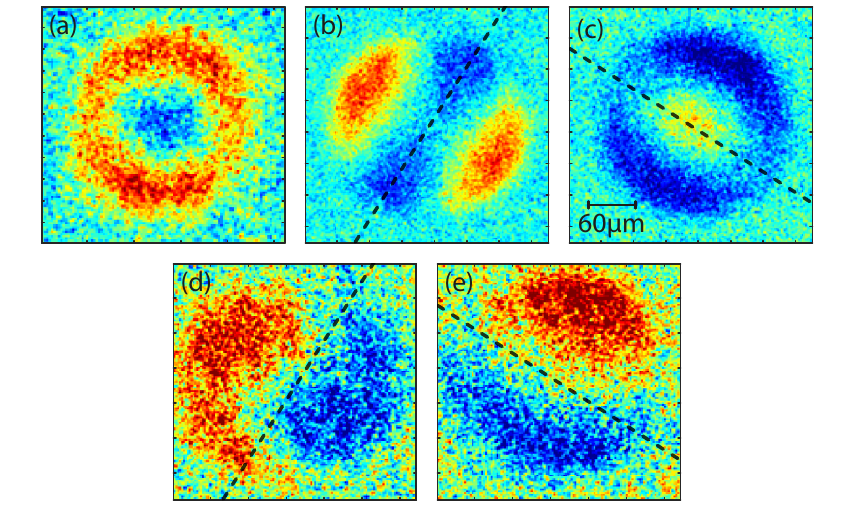}
\caption{\textbf{Principal component analysis of interacting and non-interacting datasets at a fixed magnetic field. (a):} In an interacting sample in the hydrodynamic regime the breathing mode oscillates at a single frequency independent of the anisotropy. We find only one principle component for the breathing mode. \textbf{(b,c):} In a non-interacting gas the breathing mode oscillations along the trap axes decouple and two independent principal components show up. We extract the principal trap axes from these images (dashed lines). \textbf{(d,e):} The dipole mode decouples into two principle components independent of interactions.}
\end{figure}

The oscillation frequencies of the center and width of our atom cloud are extracted by fitting a 2D Gauss function to the measured absorption images. The principal trap axes are used to fix the angle parameter of the 2D Gauss function in this fit. If the angle of the Gaussian fit does not correspond to the principal trap axis, a beat between the oscillation of both cloud center coordinates $x_0$ and $y_0$ is noticeable. This beat leads to systematic errors in the determination of the dipole frequencies $\omega_\text{D,x}$ and $\omega_\text{D,y}$. The small radial anisotropy of around $2\,\%$ prevents us from extracting the trap axes from the Gauss fit reliably. Instead, we determine them from independent measurements using a non-interacting single component gas.

For a non-interacting cloud, both the breathing and dipole mode split into two independent modes along the principal axis of the trap. We extract these independent modes by performing a principal component analysis (PCA) on the full dataset of all average images at a fixed magnetic field. A detailed explanation of the extraction of excitation modes by PCA can be found in Ref.\ \cite{Dubessy2014}. The PCA of the non-interacting gas reveals the presence of four excited modes in total, consisting of two breathing and two dipole contributions (see Fig.\ S1). When studying an interacting cloud, we find three excited modes only. In the hydrodynamic regime, the breathing mode oscillates at one single frequency. The PCA does not reveal any higher order collective excitations in either the interacting or the non-interacting gas.

We extract the principal trap axes from the principal component images of the non-interacting cloud as shown in Fig.\ S1 (b - e). The measured axes of breathing and dipole mode coincide and no dependence of the trap axis on the magnetic offset field strength $B$ is visible. With respect to our camera axis we determine the rotation angle of the principle trap axes as $(33.6\pm1.6)^\circ$. The angle parameter of the 2D Gauss fit is fixed to this value for the analysis of all our measurements.

\section{Trap Anisotropy}
After fixing the principal trap axes, the anisotropy of the harmonic confinement is extracted as
\begin{equation}
	a= \frac{\omega_\text{x}}{\omega_\text{y}}-1.
\end{equation}
In Fig.\ S2 (a) we show the measured trap anisotropies for the same interacting dataset as shown in Fig.\ 2 and 3 of the main text. We find anisotropies on the order of $1-2\,\%$ for the dataset that was taken using the mixture of hyperfine states  $\ket{1}$ and $\ket{2}$ (dark blue) and $3-4\,\%$ for the $\ket{1}$-$\ket{3}$ mixture (light blue). The difference is explained by a long downtime of the experiment between these datasets that lead to a shift of the optical trap center away from its optimal position at the center of the magnetic confinement. The breathing mode frequencies in x- and y- direction do not measure the trap anisotropy since they are locked to each other by interactions.

\begin{figure}\label{fig:s4}
\includegraphics[]{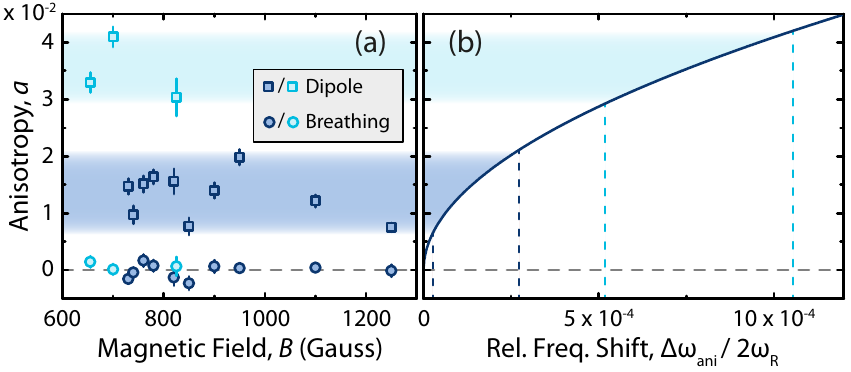}
\caption{\textbf{Characterization of the trap anisotropy. (a):} The anisotropy is extracted from the ratio of the dipole frequencies along both trap axes (squares). The frequencies of the dipole mode are locked by interactions (circles). Differently coloured measurements indicate that the measurements have been performed with different spin mixtures. \textbf{(b):} Expected shift of the relative breathing mode frequency due to trap anisotropy. The solid line displays the exact result for the frequency shift of a degenerate gas in the Thomas-Fermi limit.}
\end{figure}

The measurements of the anomalous shift for different spin mixtures shown in Fig.\ 3 reveal that the effect of the anisotropy is, to first order, negligible. The two data sets that have been taken using the $\ket{1}$-$\ket{2}$ and $\ket{1}$-$\ket{3}$ mixture (circles and squares) show no significant difference while their respective trap anisotropies differ by a factor of two. This observation can be motivated using theoretical arguments.

Applying kinetic theory, the frequency of the breathing mode of a classical gas in an anisotropic confinement in the hydrodynamic limit can be determined analytically as \cite{Merloti2013a}
\begin{equation}
\omega_\text{B,theo}^2 = \dfrac{3}{2} ( \omega_\text{x}^2+\omega_\text{y}^2 ) + \sqrt{\omega_\text{x}^2\omega_\text{y}^2 + \dfrac{9}{4} ( \omega_\text{x}^2 -\omega_\text{y}^2 ) ^2 }.
\end{equation}
This result holds equally well for a superfluid in the Thomas-Fermi limit. We define the average trap frequency as $\omega_\text{R} =1/2 ( \omega_\text{x}+\omega_\text{y} )$ and extract the expected shift of the breathing mode due to the trap anisotropy as:
\begin{equation}
\Delta \omega_\text{ani} = (\omega_\text{B,theo} - 2 \omega_\text{R}).
\end{equation}
In Fig.\ S2 (b) we show the relative frequency shift as function of the trap anisotropy $a$. Comparing to the anisotropies that have been measured in our experiment, we expect deviations of the breathing mode frequency of $0.1\,\%$ or less compared to an isotropic confinement. This is consistent with the zero temperature Monte Carlo simulation of the anomalous shift in Ref.\ \cite{Hofmann2012}. The Monte Carlo results show no significant deviations when a small anisotropy is introduced.

\section{Trap Anharmonicity}
The anharmonicity of our confinement is the explicit symmetry violation that has the largest effect on the absolute values of the frequency measurements. Studying both interacting and non-interacting samples we find that the anharmonic shifts of both breathing and dipole mode are, to first order, the same. We conclude that a comparison of the breathing mode frequency to the dipole mode frequency does already lead to the correct measurement of the anomalous frequency shift without contributions of the anharmonicity. Still, it is desirable to obtain a quantitative prediction of the effect to compare it to the anomalous shift and to identify possible further systematic frequency shifts that could be present.

The root of the anharmonicity of the radial confinement lies in the optical SWT. As explained above, the trap is created by interference of two Gaussian beams with beam waists of $w_0 \approx 600\, \mu \text{m}$. This leads to a Gaussian potential of the form $V(x) \propto - \text{Exp}[-2 x^2 /w_0^2] $. The crucial observation is that when fitting a quadratic function $f(x) = \omega_\text{opt}^2 x^2$ to this Gaussian potential, the result for $\omega_\text{opt}$ depends on the range of the fit $[-x_0,x_0]$. From a Taylor expansion of the Gauss profile, it follows that for small $x_0 < w_0$ the best fit for a harmonic confinement in the range $[-x_0,x_0]$ is approximately given by
\begin{equation}\label{eqn:anharm}
	\omega_\text{opt}\left( x_0 \right) = \omega_0 (1 - \delta \cdot x_0^2),
\end{equation}
where $\omega_0$ is the best fit in the $x_0 = 0$ limit (or the first Taylor coefficient) and $\delta$ is some constant that can in principle be determined from the Gaussian beam width $\sigma_0$.

\begin{figure}\label{fig:s7}
\includegraphics[]{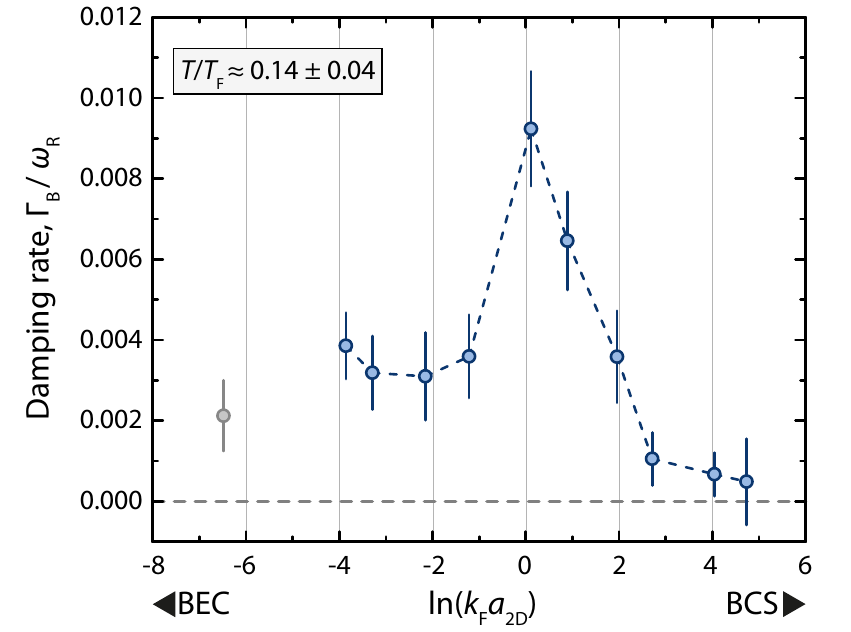}
\caption{\textbf{Damping rate measurements of the interacting breathing mode.} We measure very small damping rates $\Gamma_\text{B}$ corresponding to $1/e$-decay times on the order of 100 breathing oscillations. In the strongly interacting regime we find a significant growth of the damping rates indicating a small violation of the SO(2,1) symmetry.}
\end{figure}

Since we excite both breathing and dipole oscillation at very low amplitudes of about $8\,\%$ of the cloud width, it is reasonable to identify $x_0$ with the cloud width $\sigma$ up to some constant factor. In this picture the effect of the anharmonicity is, to first order, that the best fit for the harmonic trap frequency $\omega_\text{opt}$ for a given cloud depends quadratically on its width $\sigma$. The deviation of the actual Gaussian potential from the best harmonic fit of the form $V(x) = \omega_\text{opt}^2 x^2$ is already a higher order effect and neglected at this point. This is justified by the observation that all the measured oscillations are sinusoidal and very weakly damped (see Fig.\ S3). Non-interacting clouds show $1/e$ damping periods of more then 100 oscillations which indicates that the potential is very harmonic. Additionally, we do not resolve any further frequency shifts of the dipole mode after subtracting the first order effect of the anharmonicity.

\begin{figure}\label{fig:s2}
\includegraphics[]{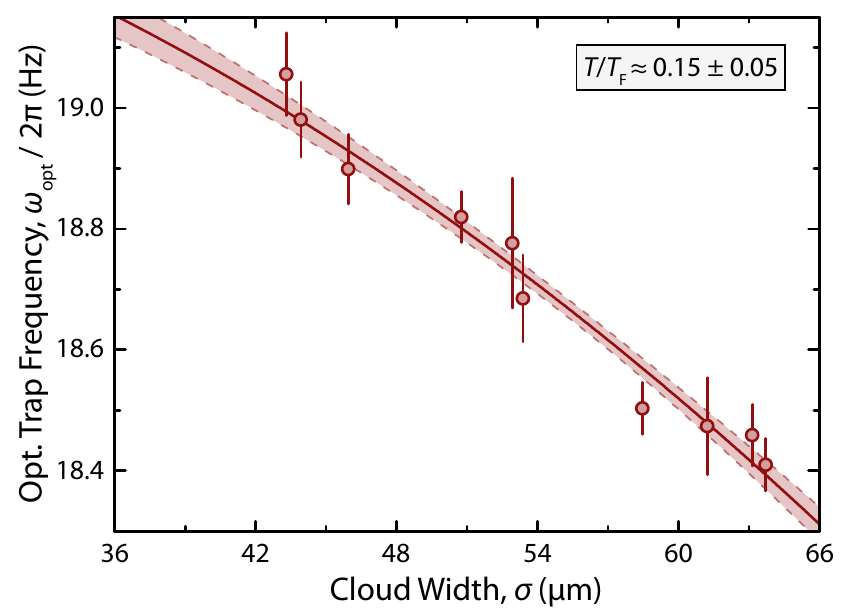}
\caption{\textbf{Frequency of the interacting dipole mode versus cloud radius.} The contributions of the magnetic confinement to the measured frequencies have been subtracted in this plot. We find the expected quadratic dependence of the frequency $\omega_\text{D}$ on the cloud radius. The solid line is a fit of equation \ref{eqn:anharm} to the data. The shaded area represents the $1\sigma$-uncertainty of the fit.}
\end{figure}

Since it is a priori not clear how the cloud width $\sigma$ and the fit range $x_0$ are related  to each other exactly and the experimentally determined value for the Gaussian width of $w_0 = 600\,\mu\text{m}$ is connected to a rather large uncertainty of around $20\,\%$, we have decided to determine $\omega_0$ and $\delta$ experimentally. We did confirm, however, that the measured correction factor $\delta$ is equal to the expected value for $w=600\,\mu\text{m}$ and $x_0 \approx 2\sigma$.

To study the anharmonic shift experimentally, we measure the dependence of the dipole mode frequency on the cloud width $\sigma$. Considering a non-interacting gas, the cloud width scales with the particle number as $\sigma \propto \sqrt[4]{N}$. Including interactions, the cloud width can in addition be controlled by the scattering length. In Fig.\ S4 the change of the frequency of the interacting dipole mode $\omega_\text{D}$ with the cloud width $\sigma$ is shown. The shown points correspond to the dataset of Fig.\ 2 and 3 of the main text. Here, the particle number has been kept fixed and the scattering length has been scanned from the BCS to the BEC limit. The contributions of the magnetic trap are subtracted from all data points. We find the expected $\sigma^2$ dependence of the optical frequency $\omega_\text{opt}$ on the cloud width and we are able to extract $\delta$ and $\omega_0$ from a fit to the data points. We observe no higher order contributions of the anharmonicity in either the interacting or the non-interacting dataset (see Fig.\ S5 b).


\begin{figure}\label{fig:s5}
\includegraphics[]{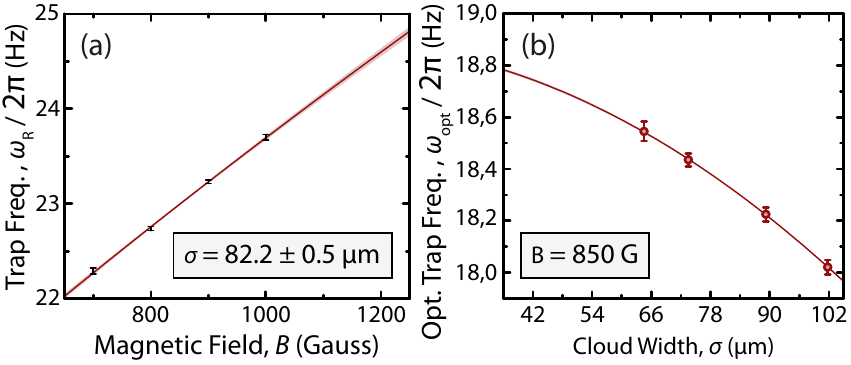}
\caption{\textbf{Trap frequency measurements with a non-interacting single component sample. (a):} By tuning the magnetic field $B$ and leaving the cloud size $\sigma$ constant, we can extract the contribution of the magnetic confinement $a\cdot B$ to the overall radial potential. The solid line is a fit of equation \ref{eqn:model1} to the data. \textbf{(b:)} The measured optical trap frequency changes with the cloud radius squared, as expected from our model. In this plot the contribution of the magnetic confinement has been subtracted from the measured values. The solid line is a fit of equation \ref{eqn:anharm} to the data.}
\end{figure}

\section{Trap Frequency Model}
We use a simple model to explain the measured trap or dipole frequencies $\omega_{R} \equiv \omega_{D}$ at any magnetic Field $B$ and cloud width $\sigma$. The total trap frequency can be estimated as
\begin{equation}\label{eqn:model1}
	\omega_\text{R}(B,\sigma) = \sqrt{\omega_\text{opt}(\sigma)^2+a\cdot B },
\end{equation}
where $\omega_\text{opt}$ is the total contribution of the optical SWT and $a\cdot B$ is the additional confinement created by the magnetic field coils. Here, $a$ is some parameter that depends on the magnetic dipole moment of the atoms and on the exact coil geometry. We fix $a$ using an independent measurement with a non-interacting gas, where we leave the cloud width $\sigma$ constant and tune only the magnetic field $B$ (see Fig.\ S5 a). To improve the fit of this model to our measurements, we include the anharmonic shifts caused by the Gaussian shaped optical potential in our model. The anharmonic shifts are especially important when studying datasets where the interactions are tuned since this leads to a large change of up to $30\,\%$ in the cloud size. Combining equations (\ref{eqn:anharm}) and (\ref{eqn:model1}) we obtain
\begin{equation}
	\omega_\text{R}(B,\sigma) = \sqrt{\omega_0^2 (1 - \delta \cdot \sigma^2)^2+a\cdot B },
\end{equation}
where $\delta = (14 \pm 1)\cdot 10^{-6}\, \mu\text{m}^{-2}$ and $\omega_0/2\pi = 19.51 \pm 0.06\, \text{Hz}$ are extracted by a fit of the model to the measured dipole frequencies (see Fig.\ S4). The model is able to explain the measured frequencies of both breathing and dipole mode for both non-interacting and interacting clouds without significant deviations. The only exception are measurements of the interacting breathing mode in the strongly interacting regime where the quantum anomaly leads to additional frequency shifts.

\section{Single Component Measurements}
In Fig.\ S5 we show trap frequency measurements using a single component gas in dependence of magnetic field $B$ (a) or cloud width $\sigma$ (b). Due to the lack of interactions scale invariance is restored in this case and the trap frequency $\omega_\text{R}$ can be extracted either from the dipole mode as $\omega_\text{R} = \omega_\text{D}$ or from the breathing mode as $\omega_\text{R} = 1/2 \omega_\text{B}$. In Fig.\ S5 we show the average of both measurements. We find that the model for the effective trap frequency $\omega_\text{R}(B,\sigma))$ explains the data very well (solid lines). No higher order systematic effects can be identified within the accuracy of our experiment.

\begin{figure}\label{fig:s6}
\includegraphics[]{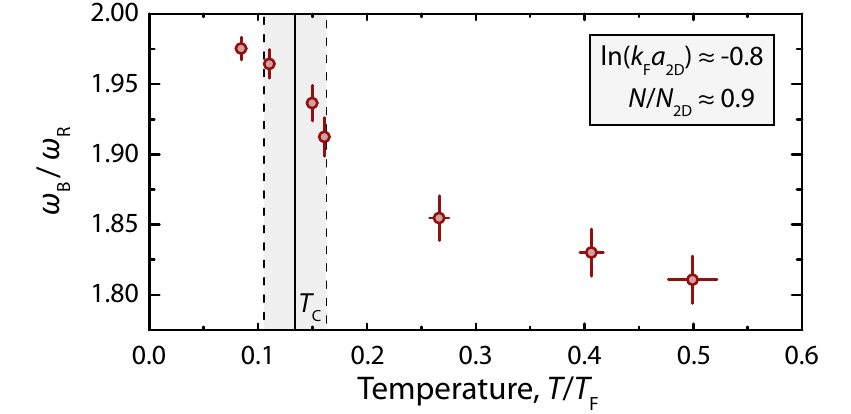}
\caption{\textbf{Temperature dependence of the breathing frequency.} The breathing mode frequency decreases quickly when the temperature of the sample rises. The strength of the 2D confinement of our SWT is given by $ \hbar \omega_\text{z}/k_\text{B} T_\text{F} \approx 0.7$.}
\end{figure}

\section{Temperature Dependence}
To control the temperature of the sample, we parametrically heat the cloud by modulating the depth of the SWT at twice the axial trap frequency $\omega_\text{z}$ with a fixed amplitude of $\approx 1\, \%$. This is followed by a hold-time of $t=600\,\text{ms}$ to let the sample reequilibrate before we excite the breathing mode oscillation. Different temperatures $T/T_\text{F}$ are reached by varying the modulation time. Due to technical limitations we can only reach higher temperatures when preparing samples with larger particle numbers, here $N/N_\text{2D} \approx 0.9$, in our current experimental setup.

We find a strong dependence of the breathing mode frequency on the temperature of the sample (see Fig.\ S6). The decrease is fastest at low temperatures and becomes slower as the temperature increases. Due to the large particle number, the frequency is in addition shifted downwards significantly by the presence of the third dimension (compare Fig.\ 4). The strong effect already at very low temperatures of $T/T_\text{F} \approx 0.1$ indicates that the measured shift is not just the result of the gas becoming more three dimensional when the temperature rises but a change of the anomalous shift itself. At the single particle level, a significant number of thermal excitations along the tightly confined axis is expected at notably higher temperatures of $T/T_F \approx 0.7$ when $k_\text{B} T = \hbar \omega_\text{z}$.

While this measurement does not allow us to completely disentangle the effects of third dimension and of temperature, it nevertheless shows that one expects that finite temperatures result in reduced anomalous shifts. This observation also explains the measurements reported by \citet{Vogt2012}, where no deviation from scale invariance has been observed at higher temperatures of $T/T_\text{F} = 0.42$ but much further in the two dimensional limit $N/N_\text{2D}\approx 0.01$. A direct quantitative comparison of this measurement to our data is difficult since we are limited to $N/N_\text{2D}\geq 0.2$ in our experiment. However, our measurements show that the expected upwards shift due to a reduced influence of the third dimension in Ref. \cite{Vogt2012} is most likely compensated by a downwards shift due to higher temperatures.

\begin{figure}\label{fig:s3}
\includegraphics[]{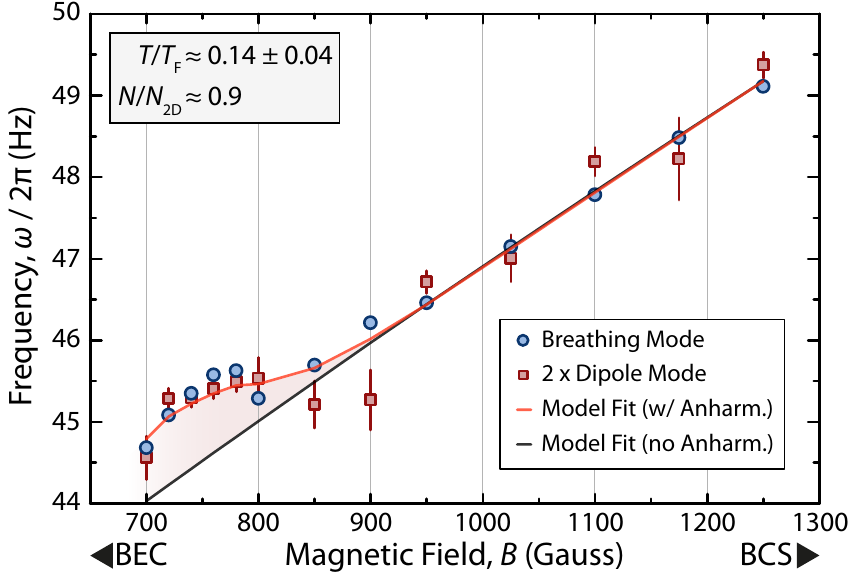}
\caption{\textbf{Measured breathing and dipole mode frequencies with larger particle numbers.} At $N/N_\text{2D}\approx 0.9$ we detect no significant violation of scale variance even at the lowest achievable temperatures in our experiment (compare Fig. 2). }
\end{figure}

In Fig.\ S7 we show the interaction dependent frequency shift of breathing and dipole mode at larger particle numbers but low temperatures. At the chosen value for $N/N_\text{2D}\approx 0.9$ scale invariance is apparently restored. From the discussion above it is clear that the symmetry is still violated but the positive anomalous frequency shift at this temperature and the negative shift due to explicit symmetry breaking of the third dimension cancel each other out.

Fig.\ S8 shows how the frequency of the breathing mode changes with interactions at a temperature of $T/T_\text{F} = 1$. As expected, the frequency tends towards the scale invariant value of $2\omega_\text{R}$ in the weakly interacting BEC and BCS limits while it is shifted far below $2\omega_\text{R}$ in the strongly interacting region. This downwards shift is a superposition of the effects of both temperature and third dimension. Negative anomalous shifts at higher temperatures are in agreement with predictions by Ref. \cite{Mulkerin2017}, where a beyond mean-field approximation shows anomalous frequency shifts of up to $-10\,\%$.

\begin{figure}\label{fig:s9}
\includegraphics[]{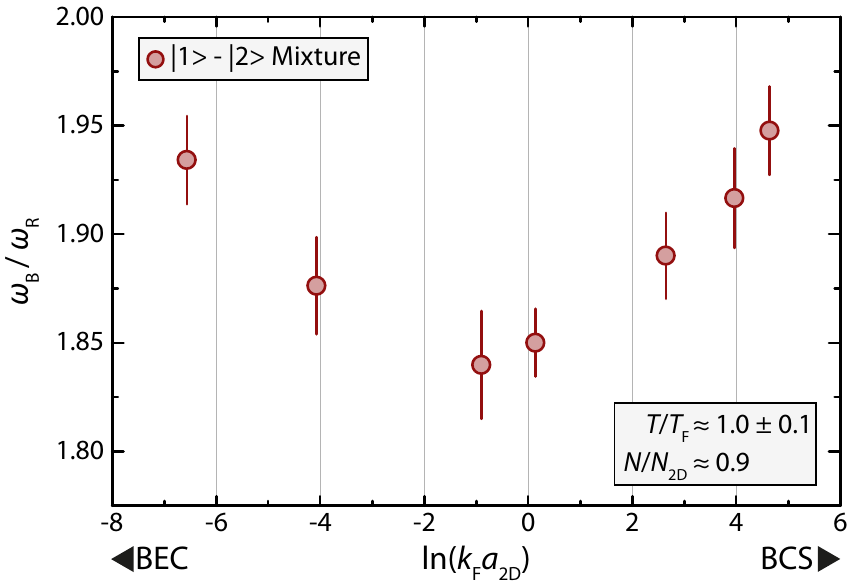}
\caption{\textbf{Frequency shift of the breathing mode at high temperatures.} The breathing mode frequency reveals a significant downwards shift at $T/T_\text{F} = 1$. The influence of both third dimension and temperature can explain this strong shift.}
\end{figure}




\providecommand{\noopsort}[1]{}\providecommand{\singleletter}[1]{#1}%

\cleardoublepage
\end{document}